\documentclass[12pt]{article}
\usepackage[scale={.8,.8},vmarginratio=1:1]{geometry}
\usepackage[latin1]{inputenc}
\usepackage{amsmath,amssymb,bbm,mathrsfs}
\usepackage[hypertex]{hyperref}

\DeclareMathOperator{\diag}{diag}

\begin{document}

\title{\vspace*{-18pt}
  {\hspace*{\fill} {\small \tt TTP07-33, SFB/CPP-07-79}}
  \\[6mm]
  \textbf{Naturalness and the Neutrino Matrix}}

\author{J. Sayre$^1$ and S. Wiesenfeldt$^{1,2}$
  \\[12pt]
  {\normalsize {\slshape
      \begin{minipage}{.75\linewidth}
        $^1$ Department of Physics, University of Illinois at
        Urbana-Champaign,
        \\
        \phantom{$^1$}\ 1110 West Green Street, Urbana, IL 61801, USA
        \\
        $^2$ Institut f\"ur Theoretische Teilchenphysik, Universit\"at
        Karlsruhe,
        \\
        \phantom{$^2$}\ 76128 Karlsruhe, Germany
      \end{minipage}
    }}
}

\date{\small }

\maketitle

\begin{abstract}
  The observed pattern of neutrino mass splittings and mixing angles
  indicates that their family structure is significantly different from
  that of the charged fermions.  We investigate the implications of
  these data for the fermion mass matrices in grand unified theories
  with a type-I seesaw mechanism.  We show that, with simple
  assumptions, naturalness leads to a strongly hierarchical Majorana
  mass matrix for heavy right-handed neutrinos and a partially cascade
  form for the Dirac neutrino matrix.  We consider various model
  building scenarios which could alter this conclusion, and discuss
  their consequences for the construction of a natural model.  We find
  that including partially lopsided matrices can aid us in generating a
  satisfying model.
\end{abstract}

\section{Introduction}\label{se:intro}

The measurement of neutrino mass splittings and mixing angles
\cite{Yao:2006px,nu-analysis} has provided a new window into physics
beyond the Standard Model.  The fact that the hierarchy between at least
one pair of the neutrinos is weak and that two leptonic mixing angles
are large, in contrast to the strongly hierarchical masses of quarks and
charged leptons and small CKM mixing, was initially surprising.  It
leads us to surmise that neutrino masses arise through a somewhat
different mechanism than the quark and charged lepton masses.  Thus, the
relation between the charged fermion and neutrino observables is not
necessarily obvious.  In fact, we have such a mechanism in the form of
the type-I seesaw \cite{Minkowski:1977sc}, which can naturally yield
neutrino masses in the range indicated by experiment.  Moreover, the
physical light neutrino mass matrix is a product of more fundamental
matrices.  This fact can potentially explain the differences between the
mixing angles and mass hierarchies of the charged fermion and neutrino
sectors.

The seesaw mechanism arises naturally within a grand-unified theory
(GUT) such as SO(10) \cite{so10}, where each generation of standard
model fermions is unified into the 16-dimensional spinor representation,
together with the right-handed neutrinos.  The breaking of $B-L$ (where
$B$ and $L$ denote baryon and lepton number, respectively), which is a
subgroup of SO(10), automatically gives rise to Majorana masses for the
singlet neutrinos, and thence to the seesaw mechanism.  Indeed, the
neutrino data have encouraged GUT model building
\cite{model,Altarelli:2004za}.

Although GUTs provide a natural framework for massive neutrinos and,
combined with family symmetries or textures, have allowed for a number
of successful models of quark masses and mixing, it has proven difficult
to incorporate neutrinos in a completely satisfactory manner.  In this
paper, we reconsider neutrino masses and mixings under the guidance of
naturalness.  That is, rather than focusing on a particular theoretical
structure and modifying it as necessary to obtain the best fit to the
data, we will try to minimize the dependence on specific model
assumptions and work up from the experimental data to see where it
naturally leads us.  In particular, we will show that the construction
of a natural, unified picture of all standard model fermion masses and
mixing angles imposes non-trivial constraints on the structure of both
sectors.

In this framework, we are interested only in the orders of magnitude of
various parameters and, in pursuing natural solutions, we seek to avoid
unnatural cancellations, i.e., that terms of a given order must cancel
to produce a term of lower order.  It may be possible to arrange such
cancellations in a technically natural way via a judicious choice of
symmetries, but this is by no means trivial.  Furthermore, an exact
symmetry is a strong assumption to make, given the current uncertainty
in the neutrino data.  We will instead adopt naturalness as described
above, seeking to constrain the approximate structure of our theory
without ad hoc symmetries.  Ideally, this structure can serve as a guide
for developing well-motivated symmetries upon which an ultimately
satisfying theory can be built.

Of course, one must make some assumptions based on previous successes to
make progress and, in this capacity, we will focus on the SO(10) models
with small representations \cite{AandB,smallrep}.  This scenario will
serve as a concrete example; however, much of the analysis could be
adapted to SO(10) models with large representations and/or type-II
seesaw mechanisms, as well as to other unifying groups.

This paper is organized as follows: We start by introducing our
theoretical framework in Section~\ref{se:theory} and reviewing the
experimental data in Section~\ref{se:exp}.  In Section~\ref{se:model} we
derive natural constraints on the neutrino mass matrices.  Since the
fermion mass matrices are related by the GUT symmetry, we study the
implications of quark mixing in Section~\ref{se:ckm}.  In
Section~\ref{se:cascade} we show how mass matrices consistent with our
constraints can be generated via family symmetries, and we investigate
how well they can fit the charged fermion masses.  In SO(10) models with
small representations, the neutrino Dirac mass matrix can receive
additional contributions via couplings to a second up-type Higgs
doublet, present in the $B-L$ breaking Higgs field.  We consider this
possibility in Section~\ref{se:new}, supplemented by an Appendix.
The remaining sections are devoted to two cases which generalize beyond
our initial assumptions. These involve models wherein otherwise
negligible leptonic rotations play an important role in neutrino mixing,
either due to a lopsided structure in some mass matrices
(Section~\ref{se:lopsided}), or to a particular form for the effective
neutrino matrix (Section~\ref{se:small}).
We conclude in Section~\ref{se:concl}.

\section{General Structure of Theory}\label{se:theory}

The standard model fermions are found in three copies of the spinor
representation $\mathbf{16}_i$.\footnote{The subscripts $i$, $j$ will be
  used to indicate generations while Higgs fields will be denoted with a
  subscript $H$.}  We will make use of the small representations
$\mathbf{10}_H$, $\mathbf{45}_H$, $\mathbf{16}_H$, $\mathbf{16}_H'$,
$\overline{\mathbf{16}}_H$, and potentially $\overline{\mathbf{16}}_H'$
to break the GUT symmetry and to generate fermion masses.  Several
authors have used this framework to build interesting models
\cite{AandB,smallrep}.

The SO(10) symmetry is broken to the Standard Model by GUT scale vacuum
expectation values (vevs), one in the SU(5) singlet direction of
$\mathbf{16}_H$ and $\overline{\mathbf{16}}_H$, denoted $v$, and
$\langle \mathbf{45}_H \rangle$ along the $B-L$ direction.  The
electroweak symmetry is broken when weak doublets in $\mathbf{10}_H$
acquire vevs.  It is also possible that the doublets in $\mathbf{16}'_H$
and $ \overline{\mathbf{16}}_H'$ acquire weak scale vevs, in which case
the light Higgs doublets are a mixture of weak doublets from the vector
and spinor representations \cite{Babu:1994kb}.  We will assume for now
that $\overline{\mathbf{16}}_H'$ does not acquire a weak vev.

Charged fermion masses are generated via several operators: the
renormalizable operator $\mathbf{16}_i \mathbf{16}_j \mathbf{10}_H$,
which contributes to all Dirac mass matrices for the standard model
fermions; the higher-dimensional operator $\mathbf{16}_i \mathbf{16}_j
\mathbf{10}_H \mathbf{45}_H$, which differentiates the quark mass
matrices from the lepton matrices due to their differing charges under
$B-L$; and $\mathbf{16}_i \mathbf{16}_j \mathbf{16}_H \mathbf{16}_H'$,
which contributes only to down quark and charged lepton mass matrices.
The operator $\mathbf{16}_i \mathbf{16}_j \mathbf{10}_H$ is symmetric in
generation space while $\mathbf{16}_i \mathbf{16}_j \mathbf{10}_H
\mathbf{45}_H$ is antisymmetric ($\mathbf{16}_i$ and $\mathbf{16}_j$ are
contracted as a $\mathbf{120}$, for $\langle 45 \rangle \propto B - L$
this is the only contraction that contributes to the mass matrices).
The operator $\mathbf{16}_i \mathbf{16}_j \mathbf{16}_H \mathbf{16}_H'$
may be symmetric or asymmetric, depending on how the fields are
contracted.

With this set of operators, the Dirac neutrino matrix $M_D$ receives
contributions from the operators $\mathbf{16}_i \mathbf{16}_j
\mathbf{10}_H$ and $\mathbf{16}_i \mathbf{16}_j \mathbf{10}_H
\mathbf{45}_H$, and we expect it to be somewhat similar to the up quark
matrix, i.e., to have a similarly strong hierarchy of mass eigenstates
from the first to the third generation. For the up quarks this is
approximately five orders of magnitude.  Although the neutrino hierarchy
can be somewhat weaker due to factors of 3 coming from the $B-L$
direction vev of the $\mathbf{45}_H$, one would still expect roughly a
$10^{-4}$ ratio between the lightest and heaviest Dirac matrix
eigenvalues.

We define the orientation of $M_D$ as $\nu^i M_D^{ij}N^j$, where $N$
is the Standard Model singlet.  Then we can parameterize the Dirac
matrix as
\begin{align}
  \label{mdirac}
  M_D \equiv L_D D_D R_D^\dagger \;.
\end{align}
Here and throughout the paper the matrices $M$ are dimensionless and the
largest eigenvalue is normalized to $1$.  Since we are primarily
concerned with interfamily relations this causes no problems, but one
should bear in mind that there is an overall scale associated with all
mass matrices.  In the above case, the dimensionful Dirac mass operator
is $u\, \nu M_D N$, where $u$ is the mass of the largest eigenvalue.
Similarly, throughout the paper $L$ and $R$ will signify unitary
matrices defined by the diagonalization equations
\begin{align}
  L^\dagger M M^\dagger L = R^\dagger M^\dagger M R = D^2 \equiv \diag
  \left( \eta^2, \epsilon^2, 1 \right) ,
\end{align} 
where $\eta$, $\epsilon$, 1 are the normalized eigenvalues of $M$.

In general, $L_D$ and $R_D$ are arbitrary unitary matrices and $D_D$ is
a diagonal matrix of the eigenvalues of $M_D$; however, we expect the
eigenvalues to be strongly hierarchical.  This hierarchy will be
naturally generated if we posit the forms
\begin{align}
  D_D \equiv \diag \left(\eta, \epsilon, 1\right) , \quad L_D \sim
  \begin{pmatrix} 
    1 & \mu' \sqrt{\frac{\eta}{\epsilon}} & \nu' \sqrt{\eta} \cr \mu'
    \sqrt{\frac{\eta}{\epsilon}} & 1 & \rho' \sqrt{\epsilon} \cr \nu'
    \sqrt{\eta} & \rho' \sqrt{\epsilon} & 1
  \end{pmatrix} 
  , \quad R_D \sim
  \begin{pmatrix} 
    1 & \mu \sqrt{\frac{\eta}{\epsilon}} & \nu \sqrt{\eta} \cr \mu
    \sqrt{\frac{\eta}{\epsilon}} & 1 & \rho \sqrt{\epsilon} \cr \nu
    \sqrt{\eta} & \rho \sqrt{\epsilon} & 1
  \end{pmatrix}
  .
  \label{dirac-structure}
\end{align}
We expect $\eta \ll \epsilon \ll 1$.  Based on the quark hierarchy we
may estimate their approximate size as $\eta \sim 10^{-4}$ and $\epsilon
\sim 10^{-2}$, but most of the analysis does not depend on this
assumption.

$L_D$ and $R_D$ are unitary matrices and the parameterizations above
should be read as giving the orders of magnitude only of the various
entries.  The parameters $\mu$, $\nu$, $\rho$ and their primed
counterparts are generally expected to be less than or equal to order
one.  If they were significantly larger, various entries would need to
cancel to preserve the smaller eigenvalues.  Thus $\mu,\, \nu,\, \rho
\sim 1$ is the minimal requirement for naturalness in the absence of an
exact symmetry relating the Yukawa couplings.  This is known as a
geometrical hierarchy pattern \cite{Dorsner:2001sg}.  It corresponds to
the following form for $M_D$:
\begin{align}
  M_D \sim 
  \begin{pmatrix}
    \leq \eta & \sqrt{\eta \epsilon} & \sqrt{\eta} \cr \sqrt{\eta
      \epsilon} & \leq \epsilon & \sqrt{\epsilon} \cr \sqrt{\eta} &
    \sqrt{\epsilon} & 1
  \end{pmatrix}
  .
\end{align}
The central feature of such a matrix is that the off-diagonal entries
play a dominant or codominant role in determining the two smaller
eigenvalues.  A geometric hierarchy can be easily obtained with a U(1)
symmetry via the Froggatt-Nielsen mechanism \cite{Froggatt:1978nt}.

On the other hand, $\mu$, $\nu$, and $\rho$ may be arbitrarily smaller
without endangering the eigenvalue hierarchy.  In this case the diagonal
entries in $M_D$ become dominant and must be correspondingly close to
the eigenvalues.  We will refer to this possibility as a sub-geometric
hierarchy.  With three generations it is, of course, possible to have a
mixed case which is partially geometric and partially sub-geometric.

There is one exception to these naturalness considerations, which occurs
if $M_D$ is highly asymmetric, i.e., if $(M_D)_{ji}$ and $(M_D)_{ij}$
are of different orders for some $i$ and $j$.  However, if it arises
only from $\mathbf{16}_i \mathbf{16}_j \mathbf{10}_H$ and $\mathbf{16}_i
\mathbf{16}_j \mathbf{10}_H \mathbf{45}_H$, we would not expect this;
these operators give symmetric and antisymmetric contributions,
respectively, which would have to be arranged to cancel in a seemingly
unnatural way.  Thus we generally expect $L_D$ and $R_D$ to have similar
values for their parameters, i.e., $\mu \sim \mu'$, $\nu \sim \nu'$ and
$\rho \sim \rho'$.

To implement the Type-I seesaw, we need a matrix for the heavy
neutrinos: $N^i M_R^{ij} N^j$.  Such a coupling may arise from
$\frac{1}{m}(M_R)_{ij}\, \mathbf{16}_i \mathbf{16}_j
\overline{\mathbf{16}}_H \overline{\mathbf{16}}_H$ when
$\overline{\mathbf{16}}_H$ acquires its GUT scale vev $v$.  This
non-renormalizable operator is suppressed by some mass $m$, which is by
default the Planck scale but which in practice may be somewhat less,
depending on the origin of the effective operator.  The seesaw formula
then gives
\begin{align}
  M_{\nu} \simeq - M_D M_R^{-1} \left(M_D\right)^T .
\end{align}
As discussed above, $M_D$, $M_\nu$ and $M_R$ are dimensionless.  The
massive parameter which sets the scale for the neutrinos is $u^2 m /
v^2$.  For $u \sim 100$ GeV, $v \sim 10^{16}$ GeV, and $m \sim
m_\text{Pl} \sim 10^{18}$ GeV, this comes out to be 0.1 eV, consistent
with the range indicated by experiment.

We stress that the discussion above depends very little on the
assumption of small representations or the vevs used to do symmetry
breaking.  One may for example use $\langle \mathbf{45}_H \rangle$
proportional to the hypercharge generator or use a $\mathbf{54}_H$ in
place of the $\mathbf{45}_H$ to accomplish the breaking from SU(5) to
the standard model \cite{small-other}.  Alternatively, we could have
used the large representation approach with $\mathbf{10}_H$,
$\mathbf{120}_H$, and $\mathbf{126}_H$, which many authors have used for
model building \cite{largerep}.  In any case, we still expect a
hierarchy in the quark and charged lepton mass matrices.  Due to SO(10)
relations, this hierarchy should manifest itself in the Dirac neutrino
matrix as well and the same naturalness considerations apply.

\section{Experimental Constraints}\label{se:exp}

The detection of neutrino oscillation is successfully explained by
massive neutrinos with non-trivial mixing.  We know two mass squared
splittings among the neutrinos and two mixing angles of the leptonic
mixing matrix, with a limit on the third for the physical light
neutrinos \cite{nu-analysis},
\begin{align}
  \tan^2 \theta_{12} & = 0.45\pm 0.05 \;; & \Delta m^2_\text{sol} &
  = (8.0 \pm 0.3) \times 10^{-5}\ \text{eV}^2 \;; \nonumber
  \\
  \sin^2 2\theta_{23} & = 1.02 \pm 0.04 \;; & \Delta m^2_\text{atm} &
  = (2.5 \pm 0.2) \times 10^{-3}\ \text{eV}^2 \;; \nonumber
  \\
  \sin^2 2 \theta_{13} & = 0 \pm 0.05 \;. &
\end{align}
Additionally, cosmological considerations place a limit on the total
mass of the neutrinos \cite{cosmicbound}, along with limits from
tritium beta decay and neutrinoless double beta decay on the electron
neutrino \cite{Yao:2006px,nu-analysis,hm}.  These experimental results
constrain the total mass of the light neutrinos to be less than or of
the order of 1~eV.  Our discussion does not depend on the exact number
since the masses are degenerate in this limit.  The bound will only
become important to our analysis if it approaches the atmospheric mass
splitting.

The mixing is characterized by the PMNS matrix, a unitary matrix
parameterized by three angles and three phases,
\begin{align}\label{pmns}
  V_\text{PMNS} & \equiv L_e^\dagger L_\nu
  \\
  & =
  \begin{pmatrix}
    c_{12}c_{13} & s_{12}c_{13} & s_{13}e^{-i\delta} \cr
    -s_{12}c_{23}-c_{12}s_{23}s_{13}e^{i\delta} &
    c_{12}c_{23}-s_{12}s_{23}s_{13}e^{i\delta} & s_{23}c_{13} \cr
    s_{12}s_{23}-c_{12}c_{23}s_{13}e^{i\delta} &
    -c_{12}s_{23}-s_{12}c_{23}s_{13}e^{i\delta} & c_{23}c_{13}
  \end{pmatrix}
  \times \diag \left( e^{i\alpha_1 /2}, e^{i\alpha_2 /2}, 1 \right) .
  \nonumber
\end{align}

For concreteness, we will assume the tribimaximal solution which sets
the mixing angles $\theta_{12} = \arcsin\left(1/\sqrt{3}\right) \simeq
35^\circ$, $\theta_{13} = 0^\circ$, $\theta_{23} = 45^\circ$
\cite{Harrison:2002er},
\begin{align}
  V_\text{PMNS} =
  \begin{pmatrix} 
    \sqrt{\frac{2}{3}} & \sqrt{\frac{1}{3}} & 0 \cr -\sqrt{\frac{1}{6}}
    & \sqrt{\frac{1}{3}} & \sqrt{\frac{1}{2}} \cr \sqrt{\frac{1}{6}} &
    -\sqrt{\frac{1}{3}} & \sqrt{\frac{1}{2}}
  \end{pmatrix}
  ,
\end{align}
neglecting phases.  This is in some sense an extreme solution consistent
with the data.  Given the several seemingly disparate factors which
influence the angles, it seems highly unlikely that any model will
predict exactly zero for $\theta_{13}$, or exactly maximal atmospheric
mixing, unless carefully designed to do so \cite{Altarelli:2004za}.
Therefore it may well be that experiments eventually favor a less
striking set of angles.  Furthermore, in a detailed model one would also
need to carefully consider renormalization, which can have a significant
effect on the mixing angles and mass splittings
\cite{Antusch:2005gp}.\footnote{For example, a bimaximal mixing scenario
  ($\theta_{12},\, \theta_{23} = 45^\circ$, $\theta_{13} = 0$) at the
  GUT scale can produce weak scale mixing angles consistent with the
  data quoted above \cite{Antusch:2005gp}.}  We do not address these
effects in further detail in this paper because they make little
difference in our analysis.  We are only looking at relative orders of
magnitude of masses and mixing angles.  Due to its simple structure, we
will use the tribimaximal solution as an experimental input.  The
critical facts we need are the existence of two large neutrino mixing
angles and a relatively weak neutrino mass hierarchy, both of which will
remain true despite renormalization effects.

We will assume for now that the tribimaximal structure is generated
essentially in the neutrino sector; given the charged lepton hierarchy,
we usually expect relatively small rotations in $L_e$ compared to the
large PMNS entries.  Since we are only concerned with orders of
magnitude, we will (for now) neglect the charged lepton component.  As
with the geometric hierarchy discussed in Section~\ref{se:theory}, there
is one exception to this rule associated with a highly asymmetric
structure, this time in the charged lepton matrix.  Such a lopsided
matrix can introduce large rotations, as shown in the Albright-Barr
model \cite{AandB}.  This case will be discussed further in
Section~\ref{se:lopsided}.

The neutrino mass matrix will be diagonalized by the tribimaximal
rotations if it has the form
\begin{align}
  M_\nu = V_\text{PMNS} D_\nu V_\text{PMNS}^T \propto &
  \begin{pmatrix}
    \left(m_1 + \frac{1}{2} m_2\right) & - \frac{1}{2}
    \left(m_1-m_2\right) & \frac{1}{2} \left(m_1-m_2\right) \cr -
    \frac{1}{2} \left(m_1-m_2\right) & \frac{1}{2}\left(\frac{1}{2} m_1
      +m_2 + \frac{3}{2} m_3 \right) & - \frac{1}{2}\left(\frac{1}{2}
      m_1 +m_2 - \frac{3}{2} m_3 \right) \cr \frac{1}{2}
    \left(m_1-m_2\right) & - \frac{1}{2}\left(\frac{1}{2} m_1 +m_2 -
      \frac{3}{2} m_3 \right) & \frac{1}{2}\left( \frac{1}{2} m_1 +m_2 +
      \frac{3}{2} m_3 \right)
  \end{pmatrix}
  ,
\end{align}
i.e., $L_\nu =R_\nu = V_{\text{PMNS}}$.  The $m$'s are the physical
neutrino masses with an arbitrary phase for $m_1$ and $m_2$.  Since we
know the two mass squared differences, we may rewrite these in terms of
a single mass,
\begin{align} 
  m_1 = e^{i\phi_1} \left|m_1\right| , \quad m_2 =
  e^{i\phi_2}\sqrt{|m_1|^2 + \Delta_\text{sol}^2} \;, \quad m_3 =
  \sqrt{|m_1|^2 + \Delta_\text{sol}^2 \pm \Delta_\text{atm}^2} \;,
\end{align}
where we have introduced the notation $\Delta\equiv\sqrt{\Delta m^2}$.
The $\pm$ in the definition of $m_3$ represents the choice of normal
$(+)$ or inverted $(-)$ hierarchy.  We take the phase factors
$e^{i\phi_{1,2}}$ to be $\pm 1$ so that there are just a few choices of
relative positive or negative to make.  Since we are only concerned with
orders of magnitude and this will give the extrema, this should not
limit the analysis.  Then it is simple to scan through the allowed range
of $m_1$.  By doing this, one can observe the patterns of relative order
in the neutrino entries which are consistent with experiment.  The
potentially interesting possibilities are
\begin{enumerate}
\item $M_\nu \sim \left( 
    \begin{smallmatrix}
      \lambda & \lambda & \lambda \cr \lambda & 1 & 1 \cr \lambda & 1 &
      1
    \end{smallmatrix} 
  \right)$, corresponding to $m_1 \ll m_2 \simeq \Delta_\text{sol}$,
  normal hierarchy.
\item $M_\nu \sim \left(
    \begin{smallmatrix} 
      0 & \lambda & \lambda \cr \lambda & 1 & 1 \cr \lambda & 1 & 1
    \end{smallmatrix} 
  \right)$, corresponding to $2m_1 \simeq m_2 \simeq \frac{2}{\sqrt{3}}
  \Delta_\text{sol}$, $\phi_2 - \phi_1 = \pi$, normal hierarchy.
\item $M_\nu \sim 
  \left(
    \begin{smallmatrix} 
      1 & 0 & 0 \cr 0 & 1 & 1 \cr 0 & 1 & 1 
    \end{smallmatrix} 
  \right)$, corresponding to $\Delta_\text{sol}(\Delta_\text{atm})
  \lesssim m_1 \simeq m_2 \lesssim \Delta_\text{atm}
  (\sqrt{2}\Delta_\text{atm})$, $\phi_2 - \phi_1 = 0$, normal (inverted)
  hierarchy.
\item $M_\nu \sim \left(
    \begin{smallmatrix} 
      1 & 0 & 0 \cr 0 & 1 & 0 \cr 0 & 0 & 1
    \end{smallmatrix}
  \right)$, corresponding to degenerate masses, $\phi_2 = 0$, $\phi_1 =
  0$.
\item $M_\nu \sim \left(
    \begin{smallmatrix} 
      1 & 0 & 0 \cr 0 & 0 & 1 \cr 0 & 1 & 0
    \end{smallmatrix} 
  \right)$, corresponding to degenerate masses, $\phi_2 = \pi$, $\phi_1
  = \pi$.
\item $M_\nu \sim \left(
    \begin{smallmatrix} 
      1 & 1 & 1 \cr 1 & 1 & 1 \cr 1 & 1 & 1
    \end{smallmatrix} 
  \right)$, corresponding to degenerate masses, $\phi_2 - \phi_1 = \pi$.
\end{enumerate}
Here $\lambda \equiv \frac{\Delta_\text{sol}}{\Delta_\text{atm}} \simeq
0.2$ and $0$ should be read as at least a few orders of magnitude
smaller than 1.  Any other possibilities should be roughly an
interpolation between those listed and we do not expect them to lead to
significant deviations from the results following.

The cases with non-degenerate masses, namely the first through third
above, violate the geometrical hierarchy naturalness limit discussed in
Section~\ref{se:theory}.  In each case the democratic 2-3 block
generically leads to two large eigenvalues of order 1 and one large
mixing angle.  Then the couplings of the first generation give a naive
estimate for the third eigenvalue of $\lambda$, $\lambda^2$, and $1$ for
the first, second, and third cases, respectively.  This is not
compatible with the eigenvalue ranges listed above, so some unexpected
cancellations would have to take place.  Moreover, these cases are more
compatible with a small $\theta_{12}$ due to the smallness of all
off-diagonal first generation entries.  The fourth and fifth cases
naturally lead to degenerate eigenvalues as listed but imply unnatural
precision to account for the large mixing angles.

In short, hierarchical neutrino masses are unexpected in conjunction
with large mixing angles, and large mixing angles naturally proceed from
large off-diagonal entries in the effective mass matrix.  Thus, case 6
above is the most natural simple assumption to account for the
experimental data; it is known as a democratic mass matrix
\cite{democratic}.

We can also consider evidence from neutrinoless double beta decay
experiments.  A positive signal would confirm the Majorana nature of
neutrinos and lend credence to seesaw models.  The experimental status
is controversial: After the Heidelberg-Moscow collaboration set the
limit $\left|m_{ee}\right| = \left|\left(M_\nu\right)_{11}\right| <
0.35\, h$ eV, where $h$ denotes the uncertainty of the nuclear matrix
element \cite{nu-analysis,hm}, a subset of the collaboration claimed
evidence for a signal \cite{KlapdorKleingrothaus:2006ff}.  Depending
on the value of $h$, this signal points at quasi-degenerate neutrino
masses in the range $0.1-0.9$ eV \cite{nu-analysis}.  This result
clearly requires confirmation from current and future experiments.  If
confirmed, the hierarchical scenarios would be ruled out, consistent
with our conclusions from naturalness.  However, since this claim is
still controversial \cite{Arnaboldi:2008ds}, we will not rule out the
hierarchical scenarios in our analysis.

We note that for an inverted hierarchy with $m_2 \simeq
\frac{3}{2\sqrt{2}}\Delta_\text{atm}$, we could have
\begin{align}
  M_\nu \sim
  \begin{pmatrix} 
    1 & 1 & 1 \cr 1 & 0 & 1 \cr 1 & 1 & 0
  \end{pmatrix} 
  , \quad
  \begin{pmatrix} 
    1  & 1 & 1 \cr 1 & 1 & 0 \cr 1 & 0 & 1
  \end{pmatrix} 
  ,
\end{align}
depending on the phases $\phi_{1,2}$.  These should be thought of as
special subcases of case 6.  As will be shown in the next section, these
possibilities will only add additional modeling constraints compared to
case 6 without additional explanatory power, so they are not
particularly interesting in this context.  Bearing these caveats in mind
we shall, however, consider some cases besides 6 because they may relax
other naturalness constraints.

\section{Modeling}\label{se:model}

Now we will do a little rearranging of the seesaw formula in terms of
the eigenvalues and unitary matrix decomposition of $M_D$:
\begin{align}
  R_D^\dagger M_R^{-1} R_D^\ast = D_D^{-1} L_D^\dagger M_\nu L_D^\ast
  D_D^{-1} .
  \label{seesaw}
\end{align}
Applying this to the sixth and henceforth canonical case above, we get
\begin{align}
  R_D^\dagger M_R^{-1} R_D^\ast \sim
  \begin{pmatrix} 
    \frac{1}{\eta^2} & \frac{1}{\eta \epsilon} & \frac{1}{\eta} \cr
    \frac{1}{\eta \epsilon} & \frac{1}{\epsilon^2} &
    \frac{1}{\epsilon} \cr \frac{1}{\eta} & \frac{1}{\epsilon} & 1
  \end{pmatrix}
  , \label{seesawdouble}
\end{align}
where we have kept only the leading terms.  The salient point is that,
with the assumption $\mu',\, \nu',\, \rho' \leq 1$, the $L_D$ rotations
(and similarly the charged lepton rotations) cannot change the orders of
the entries.  From this we see the apparent double hierarchy for $M_R$:
its eigenvalues naturally scale as $\eta^2$, $\epsilon^2$, 1 compared to
$\eta$, $\epsilon$, 1 for $M_D$.

Most of the other cases are similar and retain at least a
$\frac{1}{\eta^2}$ ratio between the first and third eigenvalues.  For
the cases where $M_\nu$ has entries less than order one, the unitary
rotations can contribute significantly, in particular they can ``fill
in" the zero entries, but they cannot make any entries larger than order
unity in $L_D^\dagger M_\nu L_D^\ast$.

There are two cases which may differ importantly from the others.  Case
1 in Section~\ref{se:exp} is interesting since it yields
\begin{align}
  \label{case1}
  R_D^\dagger M_R^{-1} R_D^\ast & \sim
  \begin{pmatrix} 
    \frac{ \lambda}{\eta^2} & \frac{1}{\eta\epsilon} \left( \lambda+
      \mu'\frac{\eta}{\epsilon}\right) & \frac{1}{\eta} \left( \lambda+
      \mu'\frac{\eta}{\epsilon}\right) \cr \frac{1}{\eta\epsilon} \left(
      \lambda + \mu'\frac{\eta}{\epsilon}\right) & \frac{1}{\epsilon^2}
    & \frac{1}{\epsilon} \cr \frac{1}{\eta} \left( \lambda+
      \mu'\frac{\eta}{\epsilon}\right) & \frac{1}{\epsilon} & 1
  \end{pmatrix}
  .
  \intertext{Similarly, for the second case we get}
  R_D^\dagger M_R^{-1} R_D^\ast & \sim
  \begin{pmatrix} 
    \frac{1}{\eta^{3/2}} \left( \lambda+
      \mu'\frac{\eta}{\epsilon}\right)
    \left(\frac{\mu'}{\sqrt{\epsilon}} + \nu'\right) & \frac{1}{\eta
      \epsilon} \left( \lambda+ \mu'\frac{\eta}{\epsilon}\right) &
    \frac{1}{\eta} \left( \lambda+ \mu'\frac{\eta}{\epsilon}\right) \cr
    \frac{1}{\eta \epsilon} \left( \lambda+
      \mu'\frac{\eta}{\epsilon}\right) & \frac{1}{\epsilon^2} &
    \frac{1}{\epsilon} \cr \frac{1}{\eta} \left( \lambda+
      \mu'\frac{\eta}{\epsilon}\right) & \frac{1}{\epsilon} & 1
  \end{pmatrix} 
  .
  \label{case2}
\end{align}
In these cases we see that we have mitigated the largest ratio of
entries from $\frac{1}{\eta^2}$ to a smaller value, although said ratio
remains significantly larger than $\frac{1}{\eta}$.

Let us now consider the effects of the matrix $R_D$ on the canonical
case.  We will show that, under the current assumptions, one can put
additional constraints on $\mu$, $\nu$ and $\rho$.  To begin, we
parameterize the inverse heavy neutrino matrix 
\begin{align}
  M_R^{-1} \equiv 
  \begin{pmatrix} 
    A & B & C \cr B & D & E \cr C & E & F
  \end{pmatrix}
  \label{eq:mrinv}
\end{align}
and evaluate both Eq.~(\ref{seesaw}) and
\begin{align}
  M_R^{-1} = R_D D_D^{-1} L_D^\dagger M_\nu L_D^\ast D_D^{-1} R_D^T
  \;,
\end{align}
which is just another rearrangement of the seesaw formula.  Keeping
only potentially leading terms, we find
{
  \begin{align}
    A &\simeq \frac{1}{\eta^2} \;, \nonumber
    \\
    B &\simeq  \frac{\mu}{\eta^{3/2} \epsilon^{1/2}} + \frac{1}{\eta
      \epsilon} \;, \nonumber
    \\[2pt]
    C &\simeq  \frac{\nu}{\eta^{3/2}} + \frac{\rho}{\eta \epsilon^{1/2}} +
    \frac{\mu \nu}{\epsilon^{3/2}} + \frac{1}{\eta} \;, \nonumber
    \\[2pt]
    D &\simeq  \frac{\mu^2}{\eta \epsilon} + \frac{2\mu}{\eta^{1/2}
      \epsilon^{3/2}} + \frac{2 \rho \mu}{\eta^{1/2}} +
    \frac{1}{\epsilon^2} \;, \nonumber
    \\[2pt]
    E &\simeq  \frac{\mu \nu}{\eta \epsilon^{1/2}} +\frac{\nu + \rho
      \mu}{\eta^{1/2} \epsilon} + \frac{\mu}{\eta^{1/2}
      \epsilon^{1/2}} + \frac{\rho \nu \epsilon^{1/2}}{\eta^{1/2}} +
    \frac{\rho}{\epsilon^{3/2}} + \frac{1}{\epsilon} \;, \nonumber
    \\[2pt]
    F &\simeq  \frac{\nu^2}{\eta} + \frac {2 \rho \nu}{\eta^{1/2}
      \epsilon^{1/2}} + \frac{2\nu}{\eta^{1/2}} +
    \frac{\rho^2}{\epsilon} + \frac{\rho}{\epsilon^{1/2}} + 1 \;.
    \label{af}
  \end{align} 
}
Now, with a little consideration, one can see that each entry should
only be as big as the rightmost term.  This is because
Eq.~(\ref{seesaw}) must still be satisfied looking only at the order of
the terms.  For example, we can look at the equation for the (12) entry
of Eq.~(\ref{seesawdouble}) in terms of $A$ through $F$ and $\mu$,
$\nu$, and $\rho$ via Eqs.~(\ref{dirac-structure}) and (\ref{eq:mrinv}).
This comes out to be
\begin{align} 
  A\, \mu \sqrt{\frac{\eta}\epsilon} + B + C \left( \rho \sqrt{\epsilon}
    + \mu \nu \frac{\eta}{\sqrt{\epsilon}} \right) + D\, \mu
  \sqrt{\frac{\eta}{\epsilon}} + E \left( \mu \rho + \nu \right)
  \sqrt{\eta} + F \rho \nu \sqrt{\eta\epsilon} \sim
  \frac{1}{\eta\epsilon} .
  \label{seesaw2}
\end{align} 
$B$ appears in this equation with a coefficient of order 1, thus any
solution to the set of conditions in Eqs.~(\ref{af}) with
$B>\frac{1}{\eta\epsilon}$ will apparently not satisfy
Eq.~(\ref{seesaw2}).\footnote{Here the important number is actually the
  ratio $B/F \sim 1/(\eta\epsilon)$.  Using the conventions above we
  find $F \sim 1$, but there is an overall numerical factor which we
  omit because it can be absorbed into the dimensionful vevs.}  This is
a naturalness condition.  One can, of course, numerically satisfy both
equations but it requires a cancellation between two terms to at least
an order of magnitude.  If we want to avoid the need for a symmetry
precisely relating various parameters, the only natural solution is to
set \mbox{$B \sim \frac{1}{\eta\epsilon}$}.\footnote{Technically, it
  could be smaller since Eq.~(\ref{seesaw}) depends on experimental
  numbers.  Thus in Eq.~(\ref{seesaw2}), it may cancel the theoretical
  parameter term $\mu/(\eta^{3/2}\sqrt{\epsilon})$ without fine tuning
  as long as it is consistent with the experimentally allowed range.  At
  any rate, it would only make the hierarchy stronger since $A \sim
  \frac{1}{\eta^2}$ regardless.}

Applying the same analysis to the rest of Eqs.~(\ref{af}), we come to
the conclusion that
\begin{align}
  M_R^{-1} \sim D_D^{-1}L_D^\dagger M_\nu L_D^\ast D_D^{-1} ,
\end{align} 
or that the hierarchy of $M_R^{-1}$ could be even stronger, regardless
of $R_D$.  Then we must impose constraints on the mixing parameters in
Eqs.~(\ref{af}) so that the parameters $B - F$ do not become too large:
\begin{align} 
  \mu \lesssim \sqrt{\frac{\eta}{\epsilon}}, \quad \nu \lesssim
  \sqrt{\eta}, \quad \rho \lesssim \sqrt{\epsilon}.
  \label{eq:constraint}
\end{align}
For $\eta \sim 10^{-4}$ and $\epsilon \sim 10^{-2}$, this corresponds
to $\mu,\,\rho \lesssim 10^{-1}$ and $\nu \lesssim 10^{-2}$.

If we take the minimum required suppression and apply it to $\mu'$,
$\nu'$, and $\rho'$ as well, we get the cascade hierarchy pattern
\cite{Dorsner:2001sg,Altarelli:1999dg} for the Dirac matrix,
\begin{align} 
  \label{cascade-def}
  M_D \sim
  \begin{pmatrix} 
    \eta & \eta & \eta \cr \eta & \epsilon & \epsilon \cr \eta &
    \epsilon & 1 
  \end{pmatrix} 
  .
\end{align} 
For any hierarchical texture of \mbox{$R_D^\dagger M_R^{-1} R_D^\ast$}
we will find that $M_R^{-1}$ generally retains the same hierarchy.
Intuitively, this is because $R_D$ will tend to smear out any hierarchy
in $M_R^{-1}$; the larger entries will be rotated into the smaller.  The
hierarchy would only be sharpened if there were a very precise relation
between $R_D$ and $M_R^{-1}$, which we have no reason to expect.  So in
general, if $R_D^\dagger M_R^{-1} R_D^\ast$ has a hierarchy of entries,
$M_R^{-1}$ should have at least as strong a hierarchy.  Conversely, to
maintain a strong hierarchy in $R_D^\dagger M_R^{-1} R_D^\ast$, the
unitary rotations cannot be too far from diagonal, a fact reflected in
the constraints on $\mu$, $\nu$ and $\rho$.

For the other possible textures of $M_\nu$ with one or two suppressed
entries, we mostly find equal or stronger constraints on the $\mu$,
$\nu$, and $\rho$.  For example, if the (23) and (32) entries of $M_\nu$
are small so that the corresponding entries in $D_D^{-1}L_D^\dagger
M_\nu L_D^\ast D_D^{-1}$ are much less than $\frac{1}{\epsilon}$, then
we also require $E \ll \frac{1}{\epsilon}$.  This in turn imposes
stronger constraints on the mixing parameters.  This is the situation
for cases 3-5 as well as the special sub-cases of 6 mentioned in
Section~\ref{se:exp}.

It is interesting that the constraints on $\mu$, $\nu$, and $\rho$
remain valid even if we take the first case of the list,
\begin{align}
  M_\nu \sim  
  \begin{pmatrix} 
    \lambda & \lambda & \lambda \cr \lambda & 1 & 1 \cr \lambda & 1 & 1
  \end{pmatrix} 
  , \qquad R_D^\dagger M_R^{-1} R_D^\ast \sim
  \begin{pmatrix} 
    \frac{\lambda }{\eta^2} & \frac{\lambda }{\eta \epsilon} &
    \frac{\lambda }{\eta} \cr \frac{\lambda }{\eta \epsilon} &
    \frac{1}{\epsilon^2} & \frac{1}{\epsilon} \cr \frac{\lambda}{\eta} &
    \frac{1}{\epsilon} & 1
  \end{pmatrix}
\end{align}
This is because we retain the strong hierarchy along the first column
and row, as well as in the (23)-block, whose entries remain less than
or equal in order to the first generation entries.

The one exceptional case is the other form noted before, case 2.  This
leads one to the conclusion
\begin{align}
  M_R^{-1}  \sim 
  \begin{pmatrix} 
    \frac{1}{\eta^{3/2}} \left( \lambda + \mu'\frac{\eta}{\epsilon}
    \right) \left (\frac{\mu'}{\sqrt{\epsilon}} + \nu' \right) &
    \frac{1}{\eta \epsilon} \left( \lambda +
      \mu'\sqrt{\frac{\eta}{\epsilon}} \right) & \frac{1}{\eta} \left(
      \lambda + \mu'\sqrt{\frac{\eta}{\epsilon}} \right) \cr
    \frac{1}{\eta \epsilon} \left( \lambda +
      \mu'\sqrt{\frac{\eta}{\epsilon}} \right) & \frac{1}{\epsilon^2} &
    \frac{1}{\epsilon} \cr \frac{1}{\eta} \left( \lambda +
      \mu'\sqrt{\frac{\eta}{\epsilon}} \right) & \frac{1}{\epsilon} & 1
  \end{pmatrix}
  ,
\end{align}
and the naturalness conditions
\begin{align}
  \label{special-constraint}
  \mu \lesssim \frac{\sqrt{\eta}}{\sqrt{\epsilon}\,\lambda} \sim 1 ,
  \qquad \nu \lesssim \frac{\sqrt{\eta}}{\lambda} \sim 0.1 , \qquad \rho
  \leq \sqrt{\epsilon} \sim 0.1 \;.
\end{align}
So in this case we are not as constrained as the cascade pattern but
still more constrained than the geometric pattern; only the constraint
on $\rho$ remains the same.  This makes sense since, in this case, we
have a relatively weak hierarchy in the first row and column compared to
the canonical case.  Therefore, we find weaker constraints on the
rotation parameters for the first generation.

In general then, we are led to both a double (or at least enhanced)
hierarchy for $M_R$ and a cascade (or sub-geometrical) pattern for $M_D$
in a simple type-I scenario.  Other authors have come to similar
conclusions following from the assumption of hierarchical Yukawa
matrices \cite{Dermisek:2004tx,Casas:2006hf}.

\section{CKM Constraints}\label{se:ckm}

The Dirac mass matrices of quarks and leptons are related by SO(10) and
possibly family symmetries.  Thus, we should also consider the size of
the unitary rotations in the up and down quark mass matrices, which are
measurable through the CKM matrix, $V_\text{CKM} \equiv L_u^\dagger
L_d$.  The experimental CKM values are \cite{Yao:2006px}
\begin{align}
  V_\text{CKM} = 
  \begin{pmatrix} 
    1 & 0.226 \pm 0.002 & [4.3 \pm 0.3] \times 10^{-3} \cr 0.23 \pm 0.01
    & 1 & [4.2 \pm 0.06] \times 10^{-2} \cr [7.4 \pm 0.8] \times 10^{-3}
    & 3.5 \times 10^{-2} & 1
  \end{pmatrix}
  .
\end{align}
If we suppose for the moment a geometric pattern for both the up and
down quark matrices, then the predicted CKM matrix is
\begin{align}
  V_\text{CKM} &\simeq 
  \begin{pmatrix} 
    1 & \sqrt{\frac{m_d}{m_s}} - \sqrt{\frac{m_u\vphantom{m_d}}{m_c}} +
    \sqrt{\frac{m_u m_s\vphantom{m_d}}{m_t m_b}} &
    \sqrt{\frac{m_d}{m_b}} - \sqrt{\frac{m_u\vphantom{m_d}}{m_t}} -
    \sqrt{\frac{m_u m_s\vphantom{m_d}}{m_c m_b}} \cr
    -\sqrt{\frac{m_d}{m_s}} + \sqrt{\frac{m_u\vphantom{m_d}}{m_c}} -
    \sqrt{\frac{m_c m_d}{m_t m_b}} & 1 & \sqrt{\frac{m_s}{m_b}} -
    \sqrt{\frac{m_c\vphantom{m_d}}{m_t}} + \sqrt{\frac{m_u m_d}{m_c
        m_b}} \cr -\sqrt{\frac{m_d}{m_b}} +
    \sqrt{\frac{m_u\vphantom{m_d}}{m_t}} - \sqrt{\frac{m_c m_d}{m_t
        m_s}}& - \sqrt{\frac{m_s}{m_b}} +
    \sqrt{\frac{m_c\vphantom{m_d}}{m_t}} - \sqrt{\frac{m_u m_d}{m_t
        m_s}}& 1
  \end{pmatrix}
  \nonumber
  \\[6pt]
  & \simeq
  \begin{pmatrix} 
    1 & 0.23 - 0.06 + 4 \times 10^{-4} & 0.03 - 0.003 - 0.008 \cr -0.23
    + 0.06 - 0.001 & 1 & 0.14 - 0.04 + 0.002 \cr -0.03 + 0.003 -0.01 &
    -0.14 + 0.04 - 7 \times 10^{-4} & 1
  \end{pmatrix}
  .
\end{align}
A few features are striking.  One is that the geometric ratio
$\sqrt{\frac{m_d}{m_s}}-\sqrt{\frac{m_u\vphantom{m_d}}{m_c}}$ nicely
reproduces the experimental value for the first-second generation mixing
\cite{Fritzsch:1977za}.  The dominant term comes from the down quark
mixing, while the contribution from the up quark mixing is significantly
too small to account for the mixing by itself.  Secondly, the down quark
contribution to the first-third mixing is too large by roughly an order
of magnitude.  Lastly, the down quark contribution to the second-third
generation mixing is also too large by roughly a factor of three.  So
the geometric hierarchy does a good job for the Cabibbo angle but gives
too much mixing with the third generation.

This result is consistent with a partially cascade structure in $M_d$
and $M_u$.\footnote{Since the largest terms come from the down quark
  sector, the CKM values are also consistent with a geometric hierarchy
  in $M_u$.  Given SO(10) relations and possible family symmetries, the
  simplest assumption is that $M_u$ has a similar hierarchy structure to
  $M_d$.}  The relatively large Cabibbo angle indicates that the down
quark matrix should be close to geometrical in the 1-2 block.  However,
it will fit the data better if it is cascade-like in the third
generation.  If the same were true of $M_D$, we would be consistent with
the second case from Section~\ref{se:exp}.  On the other hand, since
$\mathbf{16}_i \mathbf{16}_j \mathbf{16}_H \mathbf{16}_H$ only
contributes to the down quark and charged lepton matrices, the neutrino
matrix could remain completely cascade-like without conflict.

\section{Implementing the Cascade Hierarchy}\label{se:cascade}

Since we argue that a cascade texture is theoretically desirable, we
will investigate how it can be generated.  We will make use of the
Froggatt-Nielsen mechanism \cite{Froggatt:1978nt} and consider a global
$U(1) \times \mathbbm{Z}_2 \times \mathbbm{Z}_2^\prime$ symmetry.  We
introduce three SO(10) singlets $\mathbf{\phi}_i$.  The flavor symmetry
is broken spontaneously at a high scale $m$ by vevs of the singlet
fields, which we expect to be all of the same order,
$\left\langle{\phi}\right\rangle$.  The symmetry breaking is assumed to
be transmitted to quarks and leptons through interactions with heavy
particles so that the Yukawa couplings are constructed out of powers of
$\zeta \equiv \left\langle{\phi}\right\rangle/m$ with a texture dictated
by the family symmetry.

\pagebreak

We assign the following charges: 
\begin{center}
  \begin{tabular}{|c||c|c|c|c||c|c|c|}
    \hline
    Field & $\mathbf{16}_1$ & $\mathbf{16}_2$ & $\mathbf{16}_3$ &
    $\mathbf{10}_H$ & $\mathbf{\phi}_1$ & $\mathbf{\phi}_2$ &
    $\mathbf{\phi}_3$  
    \\ 
    \hline 
    \hline
    U(1) & 2 & 1 & 0 & 0 & $-1$ & 0 & 0
    \\ \hline
    $\mathbbm{Z}_2$ & $-$ & $-$ & + & + & + & $-$ & +
    \\ \hline
    $\mathbbm{Z}_2^\prime$ & $-$ & + & + & + & + & + & $-$ 
    \\ \hline
  \end{tabular}
\end{center}
Then the operator $M_{10}^{ij} \mathbf{16}_i \mathbf{16}_j
\mathbf{10}_H$ originates from $\Phi^{ij} \mathbf{16}_i \mathbf{16}_j
\mathbf{10}_H$, where $\Phi$ represents the higher-dimensional
couplings,
\begin{align}
  \Phi = 
  \begin{pmatrix}
    \frac{1}{m^4} \left(\mathbf{\phi}_1\right)^4 & \frac{1}{m^4}
    \left(\mathbf{\phi}_1\right)^3 \mathbf{\phi}_3 & \frac{1}{m^4}
    \left(\mathbf{\phi}_1\right)^2 \mathbf{\phi}_2\, \mathbf{\phi}_3 \cr
    \frac{1}{m^4} \left(\mathbf{\phi}_1\right)^3 \mathbf{\phi}_3 &
    \frac{1}{m^2} \left(\mathbf{\phi}_1\right)^2 & \frac{1}{m^2}
    \mathbf{\phi}_1\, \mathbf{\phi}_2 \cr \frac{1}{m^4}
    \left(\mathbf{\phi}_1\right)^2 \mathbf{\phi}_2\, \mathbf{\phi}_3 &
    \frac{1}{m^2} \mathbf{\phi}_1\, \mathbf{\phi}_2 & 1
  \end{pmatrix}
  ,
\end{align}
so that
\begin{align}
  M_{10} \sim
  \begin{pmatrix}
    \zeta^4 & \zeta^4 & \zeta^4 \cr \zeta^4 & \zeta^2 & \zeta^2 \cr
    \zeta^4 & \zeta^2 & 1
  \end{pmatrix}
  .
\end{align}
This is the cascade form of Eq.~(\ref{cascade-def}) with $\eta=\zeta^4$
and $\epsilon=\zeta^2$.  The same pattern can easily be reproduced in
the other operators which contribute to fermion masses. Note that in the
absence of the $\mathbbm{Z}_2$ symmetries we would have generated a
geometric hierarchy.

We must also consider whether a cascade hierarchy can naturally
accommodate the fermion masses in a unified theory.  Restricting
ourselves to two generations, the operators discussed in
Section~\ref{se:theory} contribute to the (normalized) mass matrices as
follows:
\begin{align}
  \label{twogencase}
  M_u &= 
  \begin{pmatrix}
    \alpha' & \alpha +\beta \cr \alpha - \beta & 1
  \end{pmatrix} 
  , & M_D & =
  \begin{pmatrix}
    \alpha' & \alpha -3\beta \cr \alpha + 3\beta & 1
  \end{pmatrix}
  , \nonumber
  \\
  M_d & =
  \begin{pmatrix}
    \alpha' + \gamma' & \alpha + \beta + \gamma \cr \alpha - \beta +
    \gamma & 1
  \end{pmatrix} 
  , & M_e & =
  \begin{pmatrix}
    \alpha' + \gamma' & \alpha -3\beta + \gamma \cr \alpha + 3\beta +
    \gamma & 1
  \end{pmatrix}
  .
\end{align}
Here, the terms $\alpha$ and $\alpha'$ parameterize the operator
$\mathbf{16}_i \mathbf{16}_j \mathbf{10}_H$.  The parameter $\beta$
derives from $\mathbf{16}_i \mathbf{16}_j \mathbf{10}_H \mathbf{45}_H$,
while $\gamma$ and $\gamma'$ characterize $\mathbf{16}_i \mathbf{16}_j
\mathbf{16}_H \mathbf{16}'_H$.  Looking at the determinants, we calculate
the mass ratios:
\begin{align}
  \frac{m_c}{m_t} & \simeq \left| \alpha' -\alpha^2 + \beta^2 \right| ,
  & \epsilon & \simeq \left| \alpha' -\alpha^2 + 9 \beta^2 \right| ,
  \nonumber
  \\[3pt]
  \frac{m_s}{m_b} & \simeq \left| \alpha' + \gamma' + \beta^2 -
    \left(\alpha+\gamma\right)^2 \right| , & \frac{m_\mu}{m_\tau} &
  \simeq \left| \alpha' + \gamma' + 9\beta^2 -
    \left(\alpha+\gamma\right)^2 \right| .
\end{align}
As expected, $\beta$ accounts for the difference of down quark and
charged fermion masses,
\begin{align}
  8 \beta^2 = \frac{m_\mu}{m_\tau} \mp \frac{m_s}{m_b} \simeq 
  \begin{cases}
    4 \times 10^{-2} & - \cr 8 \times 10^{-2} & +
  \end{cases}
\end{align}
where we used $\left(m_\mu/m_\tau\right)_\text{GUT} \simeq 0.06$ and
$\left(m_s/m_b\right)_\text{GUT} \simeq 0.02$.  Since we wish to
minimize off-diagonal terms in a cascade-like matrix, we will use the
smaller value for $\beta$,\footnote{The larger value, $\beta \simeq
  0.1$, leads to $\epsilon \simeq 0.1$.}
\begin{align}
  \beta \simeq 7 \times 10^{-2} \;.
\end{align}
Then we obtain
\begin{align}
  \label{eq:epsilon}
  \epsilon = \frac{m_c}{m_t} + 8 \beta^2 \simeq 7 \times 10^{-2} \;,
\end{align}
with $\left(m_c/m_t\right)_\text{GUT} \simeq 0.03$.  

In order to have a cascade form for $M_D$, we require $\alpha^\prime
\sim \alpha \pm 3 \beta \sim \epsilon$.  Since $3\beta \simeq 0.2$, this
implies $\alpha^\prime \sim 0.1$, independent of $\alpha$.  This value
of $\alpha^\prime$ can be consistent with the value of $\epsilon$ in
Eq.~(\ref{eq:epsilon}), but it needs to cancel significantly with
$\alpha^2$ to ensure a suitably small value for $m_c/m_t$.  Conversely,
$m_c/m_t$ implies $\alpha' \lesssim 10^{-2}$, which leads to a geometric
hierarchy in $M_D$.  Since we have been trying to avoid requiring the
cancellation of theoretical parameters, this simple cascade ansatz is
problematic.

One particularly attractive way out of this dilemma is to consider the
possibility that $M_D$, but not $M_u$, receives additional
contributions, e.g., via particular higher-dimensional operators.  If
such an operator gave a contribution to the (22)-element of $M_D$ of
order $\epsilon \sim 0.1$, $\alpha^\prime$ could be made sufficiently
small.  We consider such a scenario in the following section.

\section{New Contributions to $M_D$}\label{se:new}

In Section~\ref{se:model} we saw that the observed pattern of neutrino
masses and mixings leads us to an enhanced hierarchy for $M_R$, compared
to $M_D$.  One should note, however, that while $M_D$ is related to the
observed quark and charged lepton hierarchies by SO(10) and any family
symmetries, it is not directly observed.  In particular, one may include
another operator, $\mathbf{16}_i \mathbf{16}_j \overline{\mathbf{16}}_H
\overline{\mathbf{16}}_H'$.  As noted above, the weak doublet in
$\overline{\mathbf{16}}_H'$ can acquire a weak scale vev $u'$ such that
this operator potentially contributes to the up quark and neutrino
masses.  However, it can be constructed to contribute only to the Dirac
neutrino matrix.  In this case we expect $u' < u$, since $u$ is required
to generate a large top quark mass and the sum of the squares of weak
scale vevs must equal $(246\ \text{GeV})^2$.

A simple possibility for generating this operator is to integrate out
SO(10) singlets, $\mathbf{S}$, at some scale above the relevant GUT
scale vevs.  For this purpose we can propose the operators
\begin{align}
  \label{ext-op}
  \overline{M}_{ij} \mathbf{16}_i \overline{\mathbf{16}}_H \mathbf{S}_j
  + \overline{M}'_{ij} \mathbf{16}_i \overline{\mathbf{16}}_H'
  \mathbf{S}_j + m_s \left(M_S\right)_{ij} \mathbf{S}_i \mathbf{S}_j \;.
\end{align}
We assume at least three singlets to guarantee that all three
righthanded neutrinos become heavy.  As usual, we define $m_s$ to have
units of mass so that $M_S$ is dimensionless with entries of order 1 or
smaller, and similarly we normalize $\overline{M}$ and $\overline{M}'$
in Eq.~(\ref{type3}).  In the following analysis we assume that all the
$\mathbf{S}$ singlets are integrated out to generate an effective
Majorana mass for the $N$'s.  To compute this via a straightforward
seesaw mechanism, we will work in the basis where $M_S$ is diagonal and
impose the conditions
\begin{align}
  \label{ms-constraint}
  (M_S)_{ii} > \frac{v}{m_s}
\end{align}
for all $i$.

The mass matrix for the electrically neutral particles
reads\footnote{Barr calls this scenario a type-III seesaw mechanism
  \cite{Barr:2003nn}; however, it can also be understood as a product
  of two type-I mechanisms.}
\begin{align}
  \begin{array}{c}
    \begin{pmatrix}
      \nu & N & S 
    \end{pmatrix}
    \cr \cr
  \end{array}
  \begin{pmatrix}  
    0 & \frac{1}{2} u{M}_D & \frac{1}{2} u'\overline{M}' \cr \frac{1}{2}
    u{M}_D^T & 0 & \frac{1}{2} v\overline{M} \cr \frac{1}{2} u'
    \overline{M}'^T & \frac{1}{2} v \overline{M}^T & m_s M_S
  \end{pmatrix}
  \begin{pmatrix} 
    \nu \cr N \cr S 
  \end{pmatrix}
  .
\end{align}
As derived in Appendix~\ref{se:app}, the light neutrino mass matrix is
then given by
\begin{align}\label{type3}
  M_\nu \simeq M_D \left(\overline{M}^{-1}\right)^T M_S
  \overline{M}^{-1} M_D^T - \frac{x}{2} \left[ \overline{M}^\prime
    \overline{M}^{-1} M_D^T + M_D \left(\overline{M}^\prime
      \overline{M}^{-1}\right)^T \right] , \quad x \equiv
  \frac{u'v}{u\,m_s} \ll 1 \;.
\end{align}
The mass of the heaviest neutrino is of order $u^2 m_s / v^2$.  It is
crucial that, in the final formula, $M_D$ appears in all terms, i.e.,
terms quadratic in $\overline{M}^\prime \overline{M}^{-1}$ have not
appeared.

Let us study the effect of the new contributions.  We parameterize the
various matrices as follows:
\begin{align}
  {M}_R^{-1} & = \left(\overline{M}^{-1}\right)^T M_S \overline{M}^{-1}
  \equiv
  \begin{pmatrix} 
    A & B & C \cr B & D & E \cr C & E & F
  \end{pmatrix}
  , \qquad \overline{M}' \overline{M}^{-1} =
  \begin{pmatrix} 
    a & b & c \cr b' & d & e \cr c' & e' & f
  \end{pmatrix}
  ,
  \label{new-para}
\end{align}
(note that the matrix $\overline{M}^\prime \overline{M}^{-1}$ is
generally not symmetric), and
\begin{align}
  M_D^{-1} M_\nu \left({M}_D^{-1}\right)^T = {M}_R^{-1} - \frac{x}{2}
  \left[ M_D^{-1} \overline{M}^\prime \overline{M}^{-1} +
    \left(\overline{M}^\prime \overline{M}^{-1}\right)^T
    \left({M}_D^{-1}\right)^T \right] =
  \begin{pmatrix} 
    A' & B' & C' \cr B' & D' & E' \cr C' & E' & F'
  \end{pmatrix}
\end{align}
The last matrix, with primed capital letters, is the total effective
matrix which takes the place of $M_R^{-1}$ in Section~\ref{se:model}.
The unprimed capital letters parameterize the familiar heavy neutrino
matrix and the lower case letters parameterize the new terms.  Before
proceeding to consider the effects of these new terms, we note that
$M_R$ can easily acquire a double hierarchy if it is generated by
integrating out heavy singlets, as described above.  If $\overline{M}$
has a hierarchy comparable to $M_D$ and $M_S$ is roughly democratic, a
double hierarchy occurs naturally.

We can write the total effective parameters in terms of these old and
new components and perform the same analysis on the total effective
matrix ($A'-F'$) as we did on the simple type-I parameters ($A-F$) in
Section~\ref{se:model}.  Then we obtain the following set of equations:
{\allowdisplaybreaks
  \begin{align}
    A' & \simeq A + \left[\frac{a}{\eta} + b' \frac{\mu'}{\sqrt{\eta
          \epsilon}} + c' \frac{\nu'}{\sqrt{ \eta}} \right]x \nonumber
    \\
    B' & \simeq B + \frac{1}{2}\left[a \frac{\mu}{\sqrt{ \eta\epsilon}}
      + \frac{b}{\eta} + \frac{b'}{\epsilon} + c' \frac{\rho'
      }{\sqrt{\epsilon}} + d\ \frac{\mu'}{\sqrt{ \eta \epsilon}} + e'
      \frac{\nu'}{\sqrt{ \eta}} \right]x \nonumber
    \\
    C' & \simeq C + \frac{1}{2}\left[a \frac{\nu}{\sqrt{\eta}} + b'
      \frac{\rho}{\sqrt{\epsilon}} +c' + \frac{c}{\eta} + e
      \frac{\mu'}{\sqrt{ \eta\epsilon}} + f \frac{\nu'}{\sqrt{
          \eta}}\right]x \nonumber
    \\
    D' & \simeq D + \left[b \frac{\mu}{\sqrt{ \eta\epsilon}} +
      \frac{d}{\epsilon} + e' \frac{\rho'}{\sqrt{ \epsilon}}
    \right]x\nonumber
    \\
    E' & \simeq E + \frac{1}{2}\left[b \frac{\nu}{\sqrt{ \eta}} + c
      \frac{\mu}{\sqrt{ \eta\epsilon}} + d \frac{\rho }{\sqrt{
          \epsilon}} + \frac{e}{\epsilon} + e' + f \frac{\rho'}{\sqrt{
          \epsilon}}\right]x \nonumber
    \\
    F' & \simeq F + \left[c \frac{\nu}{\sqrt{ \eta}} + e
      \frac{\rho}{\sqrt{ \epsilon}} + f\right]x
    \label{eq:aprime-fprime}
  \end{align}
}
In these equations we have kept only the leading terms.  In doing so, we
make use of the important fact that the constraints on $\mu$, $\nu$, and
$\rho$ still apply.  They follow from consideration of the experimental
data and the geometric constraints on $M_D$ only.\footnote{This would
  not be the case if there were new terms in the effective total matrix
  which did not involve $M_D$.}

Although these equations still appear somewhat complicated, the
requirement that we fit the same hierarchy of orders as imposed in
Eqs.~(\ref{af}) can only be satisfied in a few ways.  In general, the
new terms give us new parameters which could play a role in a precision
fit to the data, but they will not affect the conclusions of this paper
unless they dominate over the old terms.  Let us consider the canonical
case, which implies
\begin{align}
  \begin{pmatrix} 
    A' & B' & C' \cr B' & D' & E' \cr C' & E' & F'
  \end{pmatrix}
  \propto
  \begin{pmatrix} 
    \frac{1}{\eta^2} & \frac{1}{\eta \epsilon} & \frac{1}{\eta} \cr
    \frac{1}{\eta \epsilon} & \frac{1}{\epsilon^2} & \frac{1}{\epsilon}
    \cr \frac{1}{\eta} & \frac{1}{\epsilon} & 1
  \end{pmatrix}
  .
\end{align}
Examining Eq.~(\ref{eq:aprime-fprime}), this puts some initial
constraints on our new parameters.  For example,
\begin{align}
  \frac{bx}{\eta} \leq B' \sim \frac{F'}{\eta \epsilon}.
\end{align}
These constraints may be summarized in matrix form:
\begin{align}
  \begin{pmatrix}
    a & b & c \cr b' & d & e \cr c' & e' & f
  \end{pmatrix}
  \lesssim
  \begin{pmatrix}
    \frac{1}{\eta} & \frac{1}{\epsilon} & 1 \cr \frac{1}{\eta} &
    \frac{1}{\epsilon} & 1 \cr \frac{1}{\eta} & \frac{1}{\epsilon} & 1
  \end{pmatrix}
  \frac{F'}{x} \;.
\end{align}
Taking these restrictions into account, we conclude that to satisfy
$\frac{A'}{F'} \sim \frac{1}{\eta^2}$ we must have
\begin{align}
  A \sim F' \frac{1}{\eta^2} \quad \text{or} \quad a \sim F'
  \frac{1}{\eta x} \;.
\end{align} 
The latter case is initially appealing because one can apparently trade
the strong double hierarchy constraint on $M_R$ for a weaker standard
hierarchy in $\overline{M}'\overline{M}^{-1}$ if the term involving $a$
dominates.

This turns out not to be feasible.  Recall that $\overline{M}$ and
$\overline{M}'$ have all entries of order 1 or less.  Thus, if $a = F'
\frac{1}{\eta x}$, there exists some $i$ and some $n \geq 1$ for which
\begin{align}
  \overline{M}'_{1i} = \frac{1}{n} \quad \text{and} \quad
  \overline{M}^{-1}_{i1} = \frac{n F'}{\eta x}
\end{align}
(cf.~Eq.~(\ref{new-para})).  Then the assumption that the $a$ term
dominates over $A$ gives us the inequality
\begin{align}
  \frac{1}{\eta^2} \sim \frac{a x}{\eta F'} \geq \frac{A}{F'} =
  \frac{1}{F'} \left[\left(\overline{M}^{-1}\right)^T M_S
    \overline{M}^{-1}\right]_{11} \geq \left(\frac{n}{\eta x}\right)^2
  (M_S)_{ii} F' \; ,
\end{align} 
from which we obtain $(M_S)_{ii} F' \leq \frac{x^2}{n^2}$.  Applying the
seesaw constraint on $M_S$ and inserting the definition of $x$ gives us
\begin{align}
  \frac{v}{m_s}F' \leq (M_S)_{ii} F' \leq \frac{u'^2 v^2}{n^2 u^2 m_s^2}
  \;.
\end{align}
This requires $F'$ to be too small, that is,
\begin{align}
  1 \lesssim \left[\left(\overline{M}^{-1}\right)^T M_S
    \overline{M}^{-1}\right]_{33} = F \leq F' \leq \frac{u'^2 v}{n^2 u^2
    m_s}.
\end{align}
Since $v \ll m_s$ and $u' \lesssim u$, this condition cannot be
satisfied.  Thus the additional contributions cannot dominate over the
type-I contributions or change the need for a double hierarchy.

One can instead look at case 1 from Section~\ref{se:exp}.  If the new
terms dominate in the largest ratio, which is still $A'/F'$, this
implies $\overline{M}_{i1}^{-1} = \lambda \frac{n F'}{\eta x}$.
Proceeding as in the canonical case above, one finds $\lambda (M_S)_{ii}
F' \leq \frac{x^2}{ n^2}$.  Since we require $F' \geq 1$ this is only
possible if
\begin{align}
  x^2 \geq (M_S)_{ii} \lambda n^2 \geq \frac{v}{m_s} \lambda n^2 \;,
\end{align} 
or equivalently,
\begin{align}
  \lambda \leq \frac{u'^2 v}{n^2 u^2 m_s} \ll 1 .
\end{align}
This is a very marginal case since we are relying on $v \ll m_s$ to use
the seesaw formula as a valid approximation and $\lambda \sim 0.2$.

If we proceed nonetheless, then we impose the conditions on $B'$:
\begin{align}
  \frac{\lambda}{\eta\epsilon} & \sim \frac{B'}{F'} \geq \frac{B}{F'}
  \sim \frac{1}{F'} \sum_k \overline{M}^{-1}_{k1} \overline{M}^{-1}_{k2}
  \; ,
\end{align}
which implies the constraint $\overline{M}_{i2}^{-1} \leq x / \left[ n
  \epsilon (M_S)_{ii} \right]$.  Now we turn to $D' \sim
\frac{1}{\epsilon^2}$.  By similar reasoning as in the canonical case it
can be shown that $D$ must dominate to satisfy $D'$ of the appropriate
magnitude, due to the suppression of the new terms by $x$.  Then
\begin{align}
  \frac{1}{\epsilon^2} & \sim \frac{D'}{F'} \geq \frac{D}{F'} \sim
  \frac{(M_S)_{jj}}{F'} \left(\overline{M}^{-1}_{j2}\right)^2 ,
\end{align}
for some $j \ne i$, which gives us the condition
\begin{align}
  \overline{M}^{-1}_{j2} \sim \frac{1}{\epsilon}
  \sqrt{\frac{F'}{(M_S)_{jj}}} \; .
\end{align} 
This in turn implies
\begin{align}
  \overline{M}^{-1}_{j1} \leq \frac{\lambda}{\eta}
  \sqrt{\frac{F'}{(M_S)_{jj}}} \; ,
\end{align}
so as not to violate the bound on $b$.  We find then that the new
contributions can technically dominate in the (11) entry but the type-I
terms remain comparable and dominate in other entries, still exhibiting
a strong hierarchy compared to $M_D$.

The related case 2, with $(M_\nu)_{11} \ll \lambda$, is, not
surprisingly, similar.  One finds that the new term $a$ can dominate if
$\lambda \sqrt{\frac{\eta}{\epsilon}} \leq \frac{u'^2 v}{n^2 u^2 m_s}$,
which provides somewhat more room for consistency with the seesaw
approximation.  The constraints on the matrices are
\begin{align}
  \overline{M}^{-1}_{i1} & \simeq \frac{ \lambda n F'}{x \sqrt{\eta
      \epsilon}} \;, & \overline{M}^{-1}_{i2} & \leq \frac{x}{n
    \sqrt{\eta\epsilon} \left(M_S\right)_{ii}} \;, &
  \overline{M}^{-1}_{j2} & \simeq \frac{1}{\epsilon}
  \sqrt{\frac{F'}{(M_S)_{jj}}} \;, & \overline{M}^{-1}_{j1} & \leq
  \frac{\lambda}{\eta} \sqrt{\frac{F'}{(M_S)_{jj}}} \;.
\end{align}

In both cases the new terms can dominate in some entries, but the type-I
terms remain important and retain a strong, albeit not quite double,
hierarchy. We note that this is due largely to the structure of the
theory: if $M_R$ is a dimension-five operator generated by integrating
out singlets, then a hierarchy in $\overline{M}$ similar to that in
$M_D$ naturally leads to a doubled hierarchy in $M_R$. Due to the
suppression of the new terms by $v/m_s$, $M_R$ will always play an
important role.

It is interesting that even with the $\mathbf{16}_i \mathbf{16}_j
\overline{\mathbf{16}}_H \overline{\mathbf{16}}_H'$ operator only
contributing to the neutrino sector, we still derive the cascade
constraints.  Although this operator only contributes to the Dirac
neutrino matrix, the constraints apply to the operators which generate
the up quark matrix.  This follows from the precise relations between
the higher dimensional operators induced by their common origin.  These
relations result in $M_D$ appearing in all terms of the formula for
$M_\nu$.  As a consequence of the persistent cascade constraints, we
cannot use the new terms to solve the mass splitting problems discussed
in Section~\ref{se:cascade}.

If one treats $\mathbf{16}_i \mathbf{16}_j \overline{\mathbf{16}}_H
\overline{\mathbf{16}}_H'$ and $\mathbf{16}_i \mathbf{16}_j
\overline{\mathbf{16}}_H \overline{\mathbf{16}}_H$ as independent, it is
possible to relax said constraints.  That is, in the discussion above
both operators depend on the coupling $\overline{M}_{ij} \mathbf{16}_i
\overline{\mathbf{16}}_H \mathbf{S}_j$ and are therefore related.  If we
allow them to vary arbitrarily, then the modified seesaw formula in
Eq.~(\ref{type3}) would have additional terms which did not involve
$M_D$.  In effect, we would be adding new terms to the Dirac neutrino
matrix which could strongly alter its hierarchy compared to the quarks
and charged leptons.  If this resulted in a relatively weak Dirac
neutrino hierarchy, $M_R$ would have a correspondingly weakened
hierarchy and the mixing parameter constraints would also weaken.
However, as shown above, this is not necessarily the case when one
begins with a more complete theory.

In general, if one can weaken the Dirac neutrino hierarchy without
upsetting the charged fermion hierarchies, the requirement of a double
hierarchy in $M_R$ and a cascade hierarchy in $M_D$ becomes less
restrictive, since they are specified relative to the eigenvalue
hierarchy of $M_D$.  One possibility for doing so may be to introduce a
vector-like fourth generation of down quarks and leptons at the GUT
scale.  This can relate $M_D$ to the down quark hierarchy such that the
hierarchy of $M_R$ is similar to that of the up quarks \cite{4gen}.

\section{Lopsided Models}\label{se:lopsided}

Thus far, we have not allowed for any cancellations between terms in our
equations, in keeping with our aim to eliminate unnatural models.  There
are, however, two scenarios where one must be more careful.  These are
cases where unitary rotations play a very significant role either due to
large rotations or small entries in the neutrino matrix.

In this section we will consider the first type of these cases, lopsided
models, wherein the operator $\mathbf{16}_i \mathbf{16}_j \mathbf{16}_H
\mathbf{16}_H'$ is constructed so as to contribute in a highly
asymmetrical way to the down quark and charged lepton mass matrices
\cite{AandB}.\footnote{We will not discuss the origin of these lopsided
  matrices, which, e.g., can be due to family symmetries \cite{AandB}.}
These lopsided matrices can yield a natural hierarchy while violating
the geometric pattern limit discussed above.  Lopsidedness results in
large off-diagonal terms in the unitary rotations on one side of the
matrix but not both.

To illustrate these features we will restrict ourselves to two
generations first.  The following table summarizes the three natural
cases we have discussed for a generic matrix $M$ with eigenvalues
$\epsilon$ and $1$, which is diagonalized by the unitary rotation
matrices $L$ and $R$.
\begin{center}
  \begin{tabular}{|l|c|c|c|}
    \hline
    \rule[-2pt]{0pt}{14pt}
    Hierarchy& $M$ & $L$ & $R$ \\ 
    \hline 
    \hline
    \rule[-4pt]{0pt}{20pt}
    Geometric & $\left(
      \begin{smallmatrix} 
        \epsilon & \sqrt{\epsilon} \cr \sqrt{\epsilon} & 1
      \end{smallmatrix}
    \right)$ & $\left(
      \begin{smallmatrix} 
        1 & \sqrt{\epsilon} \cr \sqrt{\epsilon} & 1
      \end{smallmatrix}
    \right)$ & $\left(
      \begin{smallmatrix} 
        1 & \sqrt{\epsilon} \cr \sqrt{\epsilon} & 1
      \end{smallmatrix}
    \right)$ 
    \\ 
    \hline
    \rule[-6pt]{0pt}{20pt}
    Cascade & $\left(
      \begin{smallmatrix} 
        \epsilon & \epsilon \cr \epsilon & 1
      \end{smallmatrix}
    \right)$ & $\left(
      \begin{smallmatrix} 
        1 & \epsilon \cr \epsilon & 1 
      \end{smallmatrix}
    \right)$ & $\left(
      \begin{smallmatrix} 
        1 & \epsilon \cr \epsilon & 1
      \end{smallmatrix}
    \right)$ 
    \\ 
    \hline
    \rule[-6pt]{0pt}{20pt}
    Lopsided & $\left(
      \begin{smallmatrix} 
        \epsilon & \epsilon \cr 1 & 1
      \end{smallmatrix}
    \right)$ & $\left(
      \begin{smallmatrix} 
        1 & \epsilon \cr \epsilon & 1
      \end{smallmatrix}
    \right)$ & $\left(
      \begin{smallmatrix} 1 & 1 \cr 1& 1
      \end{smallmatrix}
    \right)$ 
    \\
    \hline
  \end{tabular}
\end{center} 
For both the geometric and cascade cases $L$ and $R$ are similar to each
other.  As expected, the off-diagonal entries of $L$ and $R$ for the
cascade case are smaller than in the geometric case.

The lopsided case, being highly asymmetric, leads to very different
rotation matrices on the left and right.  We see that to generate large
mixing on one side, i.e., $R$ with all entries of the same order, we are
led to $L$ being closer to diagonal than in the geometric case.
Rather, it is similar to the cascade rotation matrices.  So in this
simple case, to preserve naturalness, there is a tradeoff between the
left and right sides.  If one side's unitary rotation violates the
geometric naturalness bound, the other's is concomitantly constrained to
be closer to unity.

To take potentially large mixing in the charged lepton sector into
account, we have to reevaluate our seesaw formula.  In
Eq.~(\ref{seesaw}), we neglected the rotations from the charged lepton
sector, parameterized by $L_e$ (cf.~Eq.~(\ref{pmns})).  To include them
we rewrite the formula as
\begin{align}
  R_D^\dagger M_R^{-1} R_D^\ast = D_D^{-1} V_0 M_\nu ' V_0^T D_D^{-1} ,
\end{align}
where
\begin{align}
  M_\nu ' \equiv V_\text{PMNS} D_\nu V_\text{PMNS}^T = L_e^\dagger M_\nu
  L_e^\ast, \qquad V_0 \equiv L_D^\dagger L_e \; .
\end{align}
$M_\nu '$ is the light neutrino mass matrix in the basis where the
charged leptons are diagonal.  It can have the same forms as discussed
in Section~\ref{se:exp} for $M_\nu$.  With the substitutions $M_\nu \to
M_\nu '$ and $L_D \to V_0$, the equations used above are unaltered.

The crucial difference is that the assumed form of $L_D$ in
Eq.~(\ref{dirac-structure}) does not necessarily apply to $V_0$ in the
lopsided case.  Since $V_0$ contains off-diagonal entries of order one,
we may arrange for terms of equal order to cancel each other in $V_0
M_\nu' V_0^T$.  This is not fine tuning because we are, in effect,
canceling an experimental term with a theoretical one, rather than
canceling two theoretical parameters against each other.  To put it
another way, we are simply using a theoretical term to generate an
experimental parameter of the same order.  The result is that we may be
able to have a form for $R_D^\dagger M_R^{-1} R_D^\ast $ which does not
have such a strong hierarchy, and which in turn may not imply the
restrictive cascade form for $M_D$.  In such a scenario, some or all of
the large mixing in $V_\text{PMNS}$ comes from charged lepton unitary
rotations.

To examine the lopsided case further we must see what can be said about
the matrix $V_0$. Again, we can look at the CKM matrix for possible
constraints.  It can tell us about the potential lopsidedness in the
down quark and charged lepton mass matrices. The operator we are using
to generate lopsidedness contributes to $M_e$ as the transpose of its
contribution to $M_d$, as is familiar from SU(5) models.\footnote{This
  is simply due to the fact that $\mathbf{16}_H$ breaks SO(10) to SU(5),
  so \mbox{$\mathbf{16}_i \mathbf{16}_j \mathbf{16}_H \mathbf{16}_H'$}
  is basically an SU(5) Yukawa operator for down quarks and charged
  fermions, suppressed by $v/M$.}  Hence, large rotations in $L_e$ would
coincide with large rotations in $R_d$ and vice versa.  We now see that
the experimental values are consistent with either a cascade structure
or a lopsided structure for the down quark mass matrix in the third
generation couplings.

One might hope that the relatively large 1-2 mixing, which is consistent
with a geometric hierarchy in the down quark matrix
(cf.~Section~\ref{se:ckm}), would constrain the 1-2 mixing in $R_d$.
This, however, turns out not to be the case.  We can construct a matrix
with all the desired features and generically large righthanded mixing,
e.g.,
\begin{align}
  M_d \sim
  \begin{pmatrix} 
    \frac{m_d}{m_b} & \frac{\sqrt{m_d m_s\vphantom{m^u}}}{m_b} &
    \frac{m_d}{m_b}
    \\[6pt]
    \frac{\sqrt{m_d m_s\vphantom{m^u}}}{m_b} & \frac{m_s}{m_b} &
    \frac{m_s}{m_b}
    \\[4pt]    1 & 1 & 1
  \end{pmatrix}
  , \quad L_d \sim
  \begin{pmatrix} 
    1 & \sqrt{\frac{m_d}{m_s}} & \frac{\sqrt{m_d
        m_s\vphantom{m^u}}}{m_b}
    \\[4pt]
    \sqrt{\frac{m_d}{m_s}} & 1 & \frac{m_s}{m_b}
    \\[6pt]
    \frac{\sqrt{m_d m_s\vphantom{m^u}}}{m_b} & \frac{m_s}{m_b} & 1
  \end{pmatrix}
  , \quad R_d \sim
  \begin{pmatrix} 
    1 & 1 & 1 \cr 1 & 1 & 1 \cr 1 & 1 & 1 
  \end{pmatrix}
  .
\end{align}
Thus, although the CKM matrix is highly suggestive of either a partially
cascade or lopsided form for the down quark mass matrix, it is difficult
to constrain the form of $R_d$ and its counterpart $L_e$ in the latter
case.

On the other hand, we note that since $V_\text{PMNS}$ has a small value
for the (13) entry, naturalness requires that at least one entry in the
column $L_e^{i1}$ be correspondingly small.  This suggests that we can
rule out the extreme lopsided case shown above.

Lopsidedness also modifies the eigenvalue fitting we did in
Section~\ref{se:cascade}.  Let us consider the case where $\mathbf{16}_i
\mathbf{16}_j \mathbf{16}_H \mathbf{16}_{H}'$ is lopsided and assume
that it contributes to only one off-diagonal entry in $M_d$ and $M_e$ in
a significant way.  Then the mass matrices in Eq.~(\ref{twogencase}) are
modified to
\begin{align}
  M_d & =
  \begin{pmatrix}
    \alpha' + \gamma' & \alpha + \beta \cr \alpha - \beta + \gamma & 1
  \end{pmatrix} 
  , & M_e & =
  \begin{pmatrix}
    \alpha' + \gamma' & \alpha -3\beta + \gamma \cr \alpha + 3\beta & 1
  \end{pmatrix}
  ,
\end{align}
with the corresponding eigenvalues
\begin{align}
  \frac{m_s}{m_b} & = \left| \frac{\alpha' + \gamma' + \beta^2 -
      \alpha^2 - \gamma \left(\alpha+\beta\right)}{1 +\gamma^2} \right|
  , & \frac{m_\mu}{m_\tau} & = \left| \frac{\alpha' + \gamma' + 9\beta^2
      - \alpha^2 - \gamma \left(\alpha+3\beta\right)}{1 +\gamma^2}
  \right| .
\end{align}
This yields 
\begin{align}
  \frac{m_\mu}{m_\tau} -\frac{m_s}{m_b} = \frac{2\beta \left( 4\beta -
      \gamma\right)}{1 +\gamma^2}\ \xrightarrow{\gamma \sim 1}\; \beta
  \sim 4 \times 10^{-2} \; ,
\end{align}
which is only a slight improvement over the symmetric case, a cascade
structure in $M_D$ is still inconsistent with the charged fermion
hierarchies.  Thus, in the absence of additional contributions to the
mass matrices, it seems we must rely on large, lopsided mixing between
the second and third generations to alleviate the need for a cascade
structure in the 2-3 block of $M_D$.

\section{Small Entries and Mixing}\label{se:small}

Aside from lopsided matrices, there is another scenario in which $V_0$
can play an important role.  We saw in Section~\ref{se:model} that the
entries of the first row and column of $R_D^\dagger M_R^{-1} R_D^\ast$
are smaller in the cases 1 and 2.  If we allow cancellations between
these entries of order $\lambda$ and the mixing parameter
$\mu'\sqrt{\frac{\eta}{\epsilon}}$, we might expect some qualitatively
different results.  In this more general case, we use $\mu'$, $\nu'$ and
$\rho'$ to parameterize $V_0$ rather than $L_D$.  Including the effects
of $L_e$, we no longer have the symmetry constraints $\left( \mu,\,
  \nu,\, \rho \right) \sim \left( \mu',\, \nu',\, \rho' \right)$.

We consider case 1:
\begin{align*}
  R_D^\dagger M_R^{-1} R_D^\ast \sim
  \begin{pmatrix} 
    \frac{\lambda}{\eta^2} & \frac{1}{\eta\epsilon} \left( \lambda +
      \mu'\sqrt{\frac{\eta}{\epsilon}} \right) & \frac{1}{\eta} \left(
      \lambda + \mu'\sqrt{\frac{\eta}{\epsilon}} \right) \cr
    \frac{1}{\eta\epsilon} \left( \lambda +
      \mu'\sqrt{\frac{\eta}{\epsilon}} \right) & \frac{1}{\epsilon^2} &
    \frac{1}{\epsilon} \cr \frac{1}{\eta} \left( \lambda +
      \mu'\sqrt{\frac{\eta}{\epsilon}} \right) & \frac{1}{\epsilon} & 1
  \end{pmatrix}
  .
\end{align*}
Here, although the unitary rotations remain relatively close to unity,
the rotation parameter $\mu' \sqrt{\frac{\eta}{\epsilon}}$ may be large
enough to cancel the experimental term $\lambda$.  Such cancellation is
only possible if $\mu' \sim 1$.\footnote{Since $\mu'$ includes
  contributions from $L_e$, its coefficient $\sqrt{\eta/\epsilon}$
  should be $\sqrt{m_e / m_{\mu}}$ if the charged lepton ratio is
  larger.  However, since $\sqrt{m_e / m_{\mu}} \sim 0.1 \sim
  \sqrt{\eta/\epsilon} $ under our assumptions, we keep our familiar
  notation.}  Under geometrical constraints, the (11) entry will be
$\frac{\lambda}{\eta^2}$, while the (12) entry could be much smaller
than $\frac{\lambda}{\eta \epsilon}$.  One can proceed to analyze the
mixing parameters in $R_D$ as in Section~\ref{se:model}.  Due to the
relatively large (11) entry, one finds that the constraints
\begin{align}
  \mu \lesssim \sqrt{\frac{\eta}{\epsilon}} \left( 1 +
    \frac{\mu'}{\lambda} \sqrt{\frac{\eta}{\epsilon}} \right) , \qquad
  \nu \lesssim \sqrt{\nu} \left( 1 + \frac{\mu'}{\lambda}
    \sqrt{\frac{\eta}{\epsilon}} \right) , \qquad \rho \lesssim
  \sqrt{\epsilon},
\end{align}
are required to preserve the small (12) and (13) entries.  Since this
requires $\mu \ll \mu' \sim 1$, the charged lepton rotations would have
to be significantly larger than those from the Dirac neutrino matrix, at
least for the 1-2 mixing.  This would suggest an approximately geometric
structure in the charged lepton matrix and a Dirac neutrino matrix with
very small first generation mixing.  Thus, the Dirac neutrino matrix
would have a more restricted form than the cascade hierarchy derived for
the simpler case without cancellations.  Unless some additional
information prompts us to favor these textures for $M_\nu,\ M_D$ and
$M_e$, there is no compelling reason to further pursue this route.

In the second case, where $(M_{\nu})_{11} \sim 0$, we find that the
constraints are the same as those listed in
Eq.~(\ref{special-constraint}), i.e., the same as we found for this case
without allowing for cancellations.  These results hold because,
regardless of how small $\lambda + \mu'\sqrt{\frac{\eta}{\epsilon}}$ may
be, we retain the same relative hierarchy between the first generation
entries and the same hierarchy in the 2-3 block, cf.~Eq.~(\ref{case2}).

We conclude that these potential cancellations have little effect on our
previous considerations.

\section{Outlook}\label{se:concl}

Barring cancellations or additional flavor symmetries, the observed
pattern of neutrino mass splittings and mixing angles leads us to two
related propositions for simple model building in the general context of
a grand-unified theory with type-I seesaw mechanism.  The first is a
double hierarchy, with respect to the hierarchy of the Dirac matrix,
$M_D$, in the effective heavy neutrino matrix $M_R$.  The second,
contingent upon the first, is a cascade structure in $M_D$, or a texture
which is even closer to diagonal.  These conclusions follow only from
the structure of the type-I seesaw formula, together with the
observation that the experimental neutrino data most naturally arise
from an approximately democratic effective light neutrino matrix.  If
the neutrino masses obey a normal hierarchy, i.e., \mbox{$m_1 \lesssim
  m_2 \sim \sqrt{\Delta m_{\text{sol}}^2} \ll m_3 \sim \sqrt{\Delta
    m_{\text{atm}}^2}$}, it is possible to relax these constraints, but
it remains true that $M_R$ should have an enhanced hierarchy and $M_D$
should have a sub-geometrical structure.  Moreover, in this case some
approximate symmetry must exist to generate a second large mixing angle
and a hierarchy consistent with experiment.

These conclusions are rather general and not restricted to the specific
model with small representations outlined in Section~\ref{se:theory}.
They hold for hierarchical, symmetric matrices, up to factors of order
one.  In light of the quark and charged lepton mass hierarchies, it is
natural for $M_D$ to be hierarchical.  In particular, this matrix is
closely related to the up quark matrix in many GUT models.  Family
symmetries will also tend to engender such relations.  In
Section~\ref{se:new} we showed that even adding an operator which
ostensibly only contributes to the Dirac neutrino matrix does not
necessarily relax our conclusions.

Can we implement these textures in a complete model?  We discussed a
scenario with a $U(1) \times \mathbbm{Z}_2 \times \mathbbm{Z}_2$ flavor
symmetry, where we generated a cascade structure for the Dirac matrices
through the Froggatt-Nielsen mechanism.  A double hierarchy in $M_R$ is
natural if it is an effective operator generated by integrating out
singlets coupled to $\mathbf{16}_i \overline{\mathbf{16}}_H$, where this
coupling has an eigenvalue hierarchy similar to that in $M_D$
(cf.~Ref.~\cite{Dermisek:2004tx}).  However, this structure led to
problems in the quark sector.  We have seen that the relatively large
Cabibbo angle implies that the down quark matrix is not purely
cascade-like, although a cascade structure in the third generation is
supported.  This does not necessarily conflict with a fully cascade
pattern in $M_D$, but it requires a somewhat more complicated picture
than the simple model described above.  Furthermore, in our specific
model, we rely on the antisymmetric operator $\mathbf{16}_i
\mathbf{16}_j \mathbf{10}_H \mathbf{45}_H$ to differentiate the down
quark and charged lepton matrices.  This implies that its contributions
cannot be too small.  Since it also contributes to the up-quark and
neutrino matrices, it becomes difficult to reconcile a cascade structure
in these matrices with the strong up-quark hierarchy in a natural way.

Lopsided models may provide us a way out of these potential
difficulties.  Compared with a cascade pattern, they are equally
compatible with the CKM matrix.  For the purposes of mass fitting,
lopsidedness slightly relaxes the need for large off-diagonal
contributions from $\mathbf{16}_i \mathbf{16}_j \mathbf{10}_H
\mathbf{45}_H$.  More importantly, a lopsided charged lepton matrix
introduces large rotations which contribute to the PMNS matrix.  If
these are primarily responsible for one or both of the large mixing
angles, it is possible to reduce the pull towards a double hierarchy in
$M_R$.  This in turn can relax the constraints that lead us to a cascade
structure for $M_D$ and so for $M_u$.  Exactly how much lopsidedness can
obviate the need for a double hierarchy remains an open question.  The
atmospheric mass splitting remains small compared to the quark mass
splittings, irrespective of the origin of the large mixing angles.  This
will tend to require an enhanced hierarchy in at least part of $M_R$.
Additionally, while it is technically possible that most or all of the
PMNS structure comes from charged lepton rotations, we must ask how much
can be done in a natural way.  For example, as discussed at the end of
Section~\ref{se:lopsided}, a small value for $\theta_{13}$ precludes
generically large mixing from lopsidedness in all generations.

This brings us to the nature and origin of $\theta_{13}$ in general,
which we have not addressed in detail in this paper.  We chose to leave
this an open question in light of the current uncertainty in the size of
$\theta_{13}$: only an upper bound is known.  While it is clear that the
solar and atmospheric mixing angles are large compared with those in the
quark sector, $\theta_{13}$ may or may not be comparatively small.
Actually, the experimental upper limit, approximately $10^\circ$, is of
the same order as the Cabibbo angle.  This is large enough that its
smallness compared to the other neutrino angles may be explained by
normal fluctuations of order one parameters without violating our sense
of naturalness \cite{Altarelli:2004za}.  However, if $\theta_{13}$ is
significantly closer to zero we should seek some more robust
explanation.  For the forms of $M_\nu$ listed in Section~\ref{se:exp},
this would require a symmetry closely relating various matrix elements.
Another possibility arises for partially lopsided matrices: if one large
mixing angle arises from the charged lepton sector and the other from
$M_\nu$ then it is natural to preserve a small third angle.  Clearly, it
is important to determine the order of $\theta_{13}$.

In summary, a combination of partially lopsided and partially cascade
matrices, in conjunction with an enhanced hierarchy in $M_R$, seems to
be the most natural route to explain the generic features of the quark
and lepton data in a grand-unified model.  The details of a complete
model remain to be worked out, but our conclusions follow from a fairly
general framework.  It will be interesting to see if a workable model
can be obtained with relatively simple family symmetries and what
consequences there might be for experimental predictions.

\smallskip

We would like to thank S.~Willenbrock for useful comments on the
manuscript.
This work was supported in part by the U.~S.~Department of Energy under
contract No.~DE-FG02-91ER40677, as well as the Sonderforschungsbereich
Transregio 9 Computergest\"utzte Theoretische Teilchenphysik of the
Deutsche Forschungsmeinschaft.


\begin{appendix}

\section{Derivation of Expanded Seesaw Formula}\label{se:app}

In this appendix, we derive the extended seesaw formula, given in
Eq.~(\ref{type3}).  As mentioned in Section~\ref{se:new}, it is crucial
that terms quadratic in $\overline{M}^\prime \overline{M}^{-1}$ do not
appear.  The formula was originally derived in Ref.~\cite{Barr:2003nn}
through a slightly different calculation.

\smallskip

As displayed in Eq.~(\ref{ext-op}), we propose the operators
\begin{align*} 
  W_S = \overline{M}_{ij} \mathbf{16}_i \overline{\mathbf{16}}_H
  \mathbf{S}_j + \overline{M}'_{ij} \mathbf{16}_i
  \overline{\mathbf{16}}_H' \mathbf{S}_j + m_s \left(M_S\right)_{ij}
  \mathbf{S}_i \mathbf{S}_j \;.
\end{align*}
We integrate out the singlet fields, $\mathbf{S}$, by taking a partial
derivative and setting it equal to zero,
\begin{align} 
  \frac{\partial W_S}{\partial S_j} \equiv 0: \quad &
  \mathbf{S}_i = - \frac{1}{2\,m_s} \left[ \overline{M}_{ij}
    \mathbf{16}_i \overline{\mathbf{16}}_H + \overline{M}'_{ij}
    \mathbf{16}_i \overline{\mathbf{16}}_H' \right]
  \left(M_S^{-1}\right)_{jk} \; .
\end{align}
Plugging this into our initial equation yields
\begin{align}
  W_S^\text{eff} & = - \frac{1}{4 m_s}\, \mathbf{16}_i \left[ \left(
      \overline{M} M_S^{-1} \overline{M}^T \right)_{ij}
    \overline{\mathbf{16}}_H \overline{\mathbf{16}}_H + \left(
      \overline{M}' M_S^{-1} \overline{M}'^T \right)_{ij}
    \overline{\mathbf{16}}_H' \overline{\mathbf{16}}_H' \right.
  \nonumber
  \\
  & \mspace{300mu} + \left. 2 \left( \overline{M}' M_S^{-1}
      \overline{M}^T \right)_{ij} \overline{\mathbf{16}}_H
    \overline{\mathbf{16}}_H' \right] \mathbf{16}_j \; .
\end{align}
Now we let the Higgs fields acquire their GUT and weak scale vevs and we
include the Dirac term $\nu M_D N$, where $\nu$ and $N$ are the left and
right-handed neutrinos, respectively.  Suppressing the generation
indices, we obtain
\begin{align}
  W_N & = -\frac{1}{4 m_s} \left[ v^2\, N \left( \overline{M} M_S^{-1}
      \overline{M}^T \right) N + u'^2\, \nu \left( \overline{M}'
      M_S^{-1} \overline{M}'^T \right) \nu + 2\, u'v\, \nu \left(
      \overline{M}' M_S^{-1} \overline{M}^T \right) N \right]
  \nonumber
  \\[2pt]
  & \quad\ + u\, \nu \, M_D \, N \; ,
\end{align}
We extremize with respect to $N$ and find
\begin{align}
  \frac{\partial W_N}{\partial N} \equiv 0: \quad & N = \left[
    \frac{u\, m_s}{v^2}\, M_D \left(\overline{M}^{-1}\right)^T M_S
    \overline{M}^{-1} - \frac{u'}{v}\, \overline{M}' \overline{M}^{-1}
  \right] \nu \; .
\end{align}
Inserting this into the last equation and performing a little algebra
gives the amended seesaw formula
\begin{align}
  W_\nu^\text{eff} \simeq \left\{ M_D \left(\overline{M}^{-1}\right)^T
    M_S \overline{M}^{-1} M_D^T - \frac{1}{2}\frac{u'v}{um_s} \left[
      \overline{M}' \overline{M}^{-1} M_D^T + M_D \left( \overline{M}'
        \overline{M}^{-1} \right)^T \right] \right\} \frac{u^2
    m_s}{v^2} \; .
\end{align}

\end{appendix}



\end{document}